\documentclass[11pt,twocolumn,letterpaper]{article}
\usepackage{amssymb}
\usepackage[top=1in,bottom=1in,left=1in,right=1in,columnsep=0.4in]{geometry}

\usepackage{times}
\usepackage{latexsym}
\usepackage[T1]{fontenc}
\usepackage[utf8]{inputenc}
\usepackage{microtype}
\usepackage{graphicx}
\usepackage{amsmath}
\usepackage{booktabs}
\usepackage{multirow}
\usepackage{url}
\usepackage{xcolor}
\usepackage[colorlinks=true,linkcolor=black,citecolor=blue!70!black,urlcolor=blue!70!black]{hyperref}
\usepackage{natbib}
\usepackage{titlesec}
\usepackage{caption}

\titleformat{\section}{\normalfont\large\bfseries}{\thesection}{0.5em}{}
\titleformat{\subsection}{\normalfont\normalsize\bfseries}{\thesubsection}{0.5em}{}
\titleformat{\subsubsection}{\normalfont\normalsize\bfseries}{\thesubsubsection}{0.5em}{}
\titlespacing*{\section}{0pt}{10pt plus 2pt minus 2pt}{4pt plus 1pt minus 1pt}
\titlespacing*{\subsection}{0pt}{8pt plus 2pt minus 2pt}{3pt plus 1pt minus 1pt}
\titlespacing*{\subsubsection}{0pt}{6pt plus 1pt minus 1pt}{2pt plus 1pt minus 1pt}

\renewenvironment{abstract}
  {\centerline{\textbf{Abstract}}\vspace{0.5ex}\begin{quote}\small}
  {\end{quote}\vspace{1ex}}

\title{\textbf{LLMSniffer: Detecting LLM-Generated Code via\\
GraphCodeBERT and Supervised Contrastive Learning}}

\author{
  \textbf{Mahir Labib Dihan} \quad \textbf{Abir Muhtasim} \\[0.3em]
  Department of Computer Science and Engineering \\
  Bangladesh University of Engineering and Technology (BUET) \\
  Dhaka, Bangladesh \\
  \texttt{\{mahirlabibdihan, auntor505\}@gmail.com}
}

\date{}

\begin{document}
\maketitle

\begin{abstract}
The rapid proliferation of Large Language Models (LLMs) in software development
has made distinguishing AI-generated code from human-written code a critical
challenge with implications for academic integrity, code quality assurance, and
software security. We present \textbf{LLMSniffer}, a detection framework that
fine-tunes \textbf{GraphCodeBERT} using a two-stage \textbf{supervised
contrastive learning} pipeline augmented with comment removal preprocessing and
an MLP classifier. Evaluated on two benchmark datasets---GPTSniffer and
Whodunit---LLMSniffer achieves substantial improvements over prior baselines:
accuracy increases from 70\% to 78\% on GPTSniffer (F1: 68\%$\to$78\%) and
from 91\% to 94.65\% on Whodunit (F1: 91\%$\to$94.64\%). t-SNE visualizations
confirm that contrastive fine-tuning yields well-separated, compact embeddings.
We release our model checkpoints, datasets, codes and a live interactive demo to facilitate further research.
\end{abstract}

\section{Introduction}
\label{sec:intro}

The emergence of capable code-generating LLMs---including GitHub
Copilot~\citep{copilot2021}, ChatGPT~\citep{openai2023gpt4}, and Code
Llama~\citep{roziere2023code}---has transformed how software is written.
Developers increasingly rely on these models for routine coding tasks, bug
fixes, and algorithm implementation. However, this shift introduces new
challenges: plagiarism detection systems in educational settings must account
for AI-generated submissions, code review pipelines need provenance
information, and security audits may require distinguishing model-generated
patterns from human idioms.

Educators are increasingly concerned about LLM usage in programming education
\citep{hung2024}. Existing solutions such as GPTZero or DetectGPT are primarily
tailored for general text and often struggle with code content due to the unique
structural and syntactic properties of programming languages. Detecting
LLM-generated code is fundamentally a binary classification task: given a code
snippet $x \in \mathcal{C}$, learn a function $f: \mathcal{C} \to \{0, 1\}$
indicating whether the code is human-written ($0$) or LLM-generated ($1$).

Recent efforts have focused on zero-shot detectors~\citep{shi2024,ye2024} and
ML-based detectors~\citep{idialu2024}. We emphasize the latter, building on the
GPTSniffer framework~\citep{yang2024gptsniffer}. We propose \textbf{LLMSniffer},
combining (1)~\textbf{GraphCodeBERT}~\citep{guo2021graphcodebert}, which encodes
code tokens and data-flow structure, with (2)~\textbf{supervised contrastive
learning}~\citep{khosla2020supervised}, which clusters same-class representations
while separating different-class ones, and (3) a comment-stripping preprocessing
step to remove potentially confounding stylistic metadata.

Our main contributions are:
\begin{itemize}\setlength\itemsep{0pt}
  \item A two-stage contrastive fine-tuning strategy for GraphCodeBERT with
        comment removal preprocessing and an MLP classification head, achieving
        state-of-the-art results on two benchmarks.
  \item Detailed ablation analysis comparing a linear classifier baseline
        against our improved MLP-based architecture.
  \item Public model checkpoints, datasets on Hugging Face, an interactive
        Streamlit demo, and a video demonstration for full reproducibility.
\end{itemize}

\section{Related Work}
\label{sec:related}

\noindent\textbf{AI-Generated Text Detection.}
Detecting machine-generated text has received considerable attention since
GPT-2~\citep{radford2019language}. Early approaches used statistical signals
such as perplexity~\citep{gehrmann2019gltr} and likelihood
ratios~\citep{mitchell2023detectgpt}, while more recent work focuses on
fine-tuned classifiers~\citep{solaiman2019release} and
watermarking~\citep{kirchenbauer2023watermark}. These methods target natural
language and do not transfer directly to source code due to syntactic and
structural differences.

\smallskip
\noindent\textbf{LLM-Generated Code Detection.}
\citet{yang2024gptsniffer} introduced \textbf{GPTSniffer}, one of the first
dedicated benchmarks for detecting ChatGPT-generated code, using a CodeBERT
encoder trained with binary cross-entropy loss. \citet{nguyen2024whodunit}
presented \textbf{Whodunit}, focusing on GPT-4-generated Python solutions to
CodeChef problems and demonstrating that code-specific features substantially
improve detection. \citet{idialu2024} further explored zero-shot and ML-based
pipelines on competitive programming benchmarks. Our work builds directly on
these benchmarks and surpasses both baselines.

\smallskip
\noindent\textbf{Code Representation Learning.}
Pre-trained models such as CodeBERT~\citep{feng2020codebert},
GraphCodeBERT~\citep{guo2021graphcodebert}, and CodeT5~\citep{wang2021codet5}
advance code understanding by encoding token sequences alongside structural
information (ASTs, data-flow graphs). We leverage GraphCodeBERT for its
demonstrated superiority on code search and clone detection.

\smallskip
\noindent\textbf{Contrastive Learning.}
Supervised contrastive learning~\citep{khosla2020supervised} extends
self-supervised contrastive objectives~\citep{chen2020simclr} to the labeled
setting. It has been applied to NLP fine-tuning~\citep{gunel2021supervised}
and code clone detection~\citep{ding2022ccclr}, but not, to our knowledge,
to AI-generated code detection prior to this work.

\section{Background}
\label{sec:background}

\subsection{GraphCodeBERT}

GraphCodeBERT~\citep{guo2021graphcodebert} is a transformer-based model
pre-trained on code--documentation pairs from
CodeSearchNet~\citep{husain2019codesearchnet}. Unlike CodeBERT, it augments the
self-attention mechanism with a data-flow graph that captures variable-level
semantic dependencies (e.g., \textit{where a variable is computed} and
\textit{where it is used}). Two additional pre-training objectives---edge
prediction and node alignment---encourage the model to internalize these
structural relationships. We use the \texttt{[CLS]} token representation as a
fixed-dimensional embedding $\mathbf{h} \in \mathbb{R}^{768}$ for downstream classification.

\subsection{Supervised Contrastive Learning}

Given a batch of $N$ labeled pairs $\{(\mathbf{x}_i, y_i)\}_{i=1}^{N}$, the
supervised contrastive loss~\citep{khosla2020supervised} is:
\begin{equation}
\mathcal{L}_{\mathrm{SupCon}} = \sum_{i=1}^{N} \frac{-1}{|P(i)|}
  \sum_{p \in P(i)} \log \frac{e^{\mathbf{z}_i \cdot \mathbf{z}_p / \tau}}
  {\displaystyle\sum_{a \in A(i)} e^{\mathbf{z}_i \cdot \mathbf{z}_a / \tau}}
\label{eq:supcon}
\end{equation}
where $\mathbf{z}_i = \ell_2\text{-normalize}(\mathrm{MLP}(\mathbf{h}_i))$ is the
projected representation, $P(i) = \{p \neq i : y_p = y_i\}$ is the set of
same-class positives, $A(i) = \{a \neq i\}$ is all other examples, and $\tau$
is a temperature hyperparameter. This objective directly optimizes for inter-class
separation and intra-class compactness in embedding space.

\section{Methodology}
\label{sec:method}

\begin{figure*}[t]
  \centering
  \includegraphics[width=\textwidth]{overview.png}
  \caption{Overview of LLMSniffer. Stage~1 trains the encoder and projection
  head with supervised contrastive loss. Stage~2 freezes the encoder and trains
  an MLP classifier with binary cross-entropy loss.}
  \label{fig:overview}
\end{figure*}

\subsection{Preprocessing: Comment Removal}

A key design decision in LLMSniffer is removing code comments before encoding.
LLM-generated code often includes verbose, explanatory comments that reflect
LLM writing style rather than algorithmic structure. Stripping comments ensures
the model learns to distinguish based on syntactic and structural code properties,
reducing the risk of exploiting superficial stylistic cues and improving
generalization.

\subsection{Model Architecture}

LLMSniffer consists of three components (Figure~\ref{fig:overview}):

\smallskip
\noindent\textbf{(1) GraphCodeBERT Encoder.} After comment removal, the code
snippet is tokenized and fed along with its data-flow graph edges into
GraphCodeBERT. The \texttt{[CLS]} embedding $\mathbf{h} \in \mathbb{R}^{768}$
serves as the code representation.

\smallskip
\noindent\textbf{(2) Projection Head.} A two-layer MLP with ReLU activation maps
$\mathbf{h}$ to $\mathbf{z} \in \mathbb{R}^{128}$ for contrastive learning. The
head is discarded at inference time, following~\citet{chen2020simclr}.

\smallskip
\noindent\textbf{(3) MLP Classification Head.} A multi-layer perceptron with
sigmoid activation maps $\mathbf{h}$ to a binary prediction (human-written or
AI-generated). We find that this MLP head outperforms the simpler linear
classifier used in prior work.

\subsection{Training Procedure}

Training proceeds in two stages.

\smallskip
\noindent\textbf{Stage~1: Contrastive Pre-training.} We fine-tune the encoder
and projection head using $\mathcal{L}_{\mathrm{SupCon}}$
(Equation~\ref{eq:supcon}) with temperature $\tau{=}0.07$. Batches are
constructed with balanced classes to maximize the number of positive pairs
per anchor.

\smallskip
\noindent\textbf{Stage~2: Classification Fine-tuning.} The projection head is
discarded, the MLP classification head is attached, and the full model is
fine-tuned end-to-end with binary cross-entropy loss. The contrastive
pre-training step warms up the encoder to produce discriminative representations
before the classifier is trained, following the approach of \citet{khosla2020supervised}.

\subsection{Implementation Details}

We initialize from the \texttt{microsoft/graphcodebert-base} checkpoint
(\texttt{HuggingFace}). Both training stages use AdamW~\citep{loshchilov2019decoupled}
with learning rate $2{\times}10^{-5}$ and linear warmup over 10\% of training
steps. Maximum sequence length is 512 tokens. Batch sizes are 32 for contrastive
training and 64 for classification fine-tuning. The projection head uses hidden
dimension 512 and output dimension 128. All experiments are conducted on a
single GPU.

\section{Experimental Setup}
\label{sec:setup}

\subsection{Datasets}

\noindent\textbf{GPTSniffer}~\citep{yang2024gptsniffer} is a dataset of code
snippets with binary labels indicating human- or ChatGPT-authorship, spanning
multiple programming languages and competitive programming problem types.
We use the official split: 600 human + 600 ChatGPT samples for training, and
131 human + 142 ChatGPT samples for testing. We use the dataset hosted at
\url{https://huggingface.co/datasets/mahirlabibdihan/GPTSniffer}.
Crucially, the test set comes from a \emph{completely different domain}
than the training set, making this a challenging cross-domain evaluation.

\smallskip
\noindent\textbf{Whodunit}~\citep{nguyen2024whodunit} is a dataset of Python
solutions to CodeChef problems, with code generated by GPT-4 and human-written
counterparts. We use: 639 human + 639 ChatGPT samples for training, and
159 human + 159 ChatGPT samples for testing. The dataset is hosted at
\url{https://huggingface.co/datasets/mahirlabibdihan/Whodunit}.

\smallskip
\noindent\textbf{Pre-training Data.} Both datasets are augmented with
CodeSearchNet~\citep{husain2019codesearchnet} (human-written code) and
CodeNet~\citep{codenet2021} (human competitive programming solutions) for
encoder warming.

\subsection{Baselines and Ablations}

We compare against two baselines and an intermediate ablation:
\begin{itemize}\setlength\itemsep{0pt}
  \item \textbf{GPTSniffer}~\citep{yang2024gptsniffer}: CodeBERT fine-tuned with
        binary cross-entropy loss.
  \item \textbf{Whodunit}~\citep{nguyen2024whodunit}: Feature-based XGBoost
        classifier with 140 hand-crafted code features.
  \item \textbf{LLMSniffer (Linear)}: Our Stage 1+2 pipeline with a linear
        classification head instead of an MLP (no comment removal).
  \item \textbf{LLMSniffer (Ours)}: Full pipeline with comment removal and MLP
        classification head.
\end{itemize}

\subsection{Evaluation Metrics}

We report Accuracy, Precision, Recall, and F1 Score on each benchmark's test
set. All metrics are macro-averaged across the two classes.

\section{Results}
\label{sec:results}

\subsection{GPTSniffer Benchmark}

Table~\ref{tab:gptsniffer} compares all systems on the GPTSniffer test set.

\begin{table}[t]
\centering
\small
\setlength{\tabcolsep}{5pt}
\begin{tabular}{lcccc}
\toprule
\textbf{Model} & \textbf{Acc.} & \textbf{Prec.} & \textbf{Rec.} & \textbf{F1} \\
\midrule
GPTSniffer~\citeyearpar{yang2024gptsniffer}   & 0.70 & 0.77 & 0.69 & 0.68 \\
LLMSniffer (Linear) & 0.70 & 0.81 & 0.69 & 0.67 \\
\textbf{LLMSniffer (Ours)} & \textbf{0.78} & \textbf{0.78} & \textbf{0.77} & \textbf{0.78} \\
\bottomrule
\end{tabular}
\caption{Results on the GPTSniffer test set (macro-averaged). Metrics are
reported as proportions. The linear classifier variant shows no improvement
over the baseline; the MLP with comment removal achieves +8 pp accuracy and
+10 pp F1.}
\label{tab:gptsniffer}
\end{table}

The linear classifier variant fails to improve over the GPTSniffer baseline,
achieving the same overall accuracy (0.70). This highlights that contrastive
pre-training alone is insufficient---the classification head architecture
and comment removal are both important components. Our full model (MLP + comment
removal) achieves 0.78 accuracy and 0.78 F1, improving both precision and
recall simultaneously, unlike the prior baseline which suffered from a severe
precision--recall imbalance (high precision, low recall for the human class).

\subsection{Whodunit Benchmark}

Table~\ref{tab:whodunit} presents results on the Whodunit benchmark.

\begin{table}[t]
\centering
\small
\setlength{\tabcolsep}{5pt}
\begin{tabular}{lcccc}
\toprule
\textbf{Model} & \textbf{Acc.} & \textbf{Prec.} & \textbf{Rec.} & \textbf{F1} \\
\midrule
Whodunit~\citeyearpar{nguyen2024whodunit} & 0.91 & 0.91 & 0.91 & 0.91 \\
\textbf{LLMSniffer} & \textbf{0.947} & \textbf{0.949} & \textbf{0.943} & \textbf{0.946} \\
\midrule
$\Delta$ & +3.65 & +3.94 & +3.34 & +3.64 \\
\bottomrule
\end{tabular}
\caption{Results on the Whodunit test set. Consistent gains across all metrics.}
\label{tab:whodunit}
\end{table}

On Whodunit, LLMSniffer achieves consistent improvements across all metrics
(+3.6 pp), reflecting the dataset's balanced structure and the robustness
of our approach to GPT-4 generated code in Python.

\subsection{Embedding Visualization}

\begin{figure*}[t]
  \centering
  \includegraphics[width=0.49\textwidth]{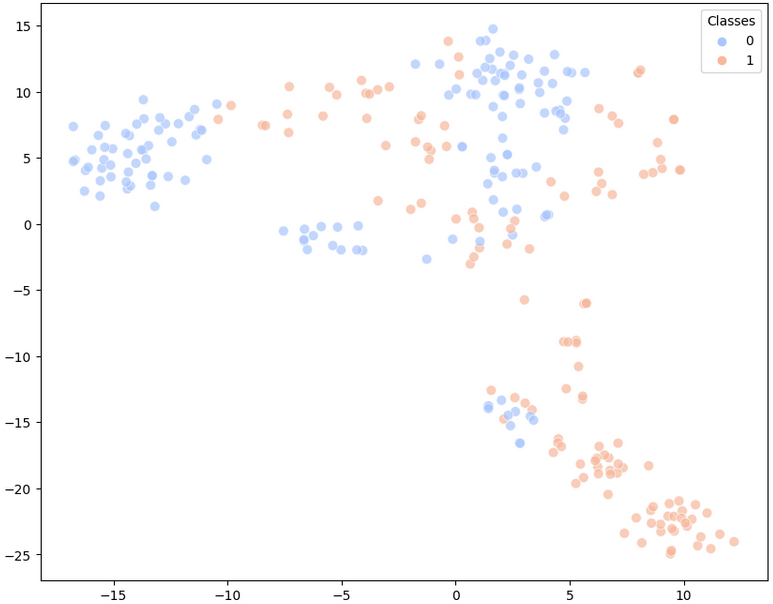}
  \hfill
  \includegraphics[width=0.49\textwidth]{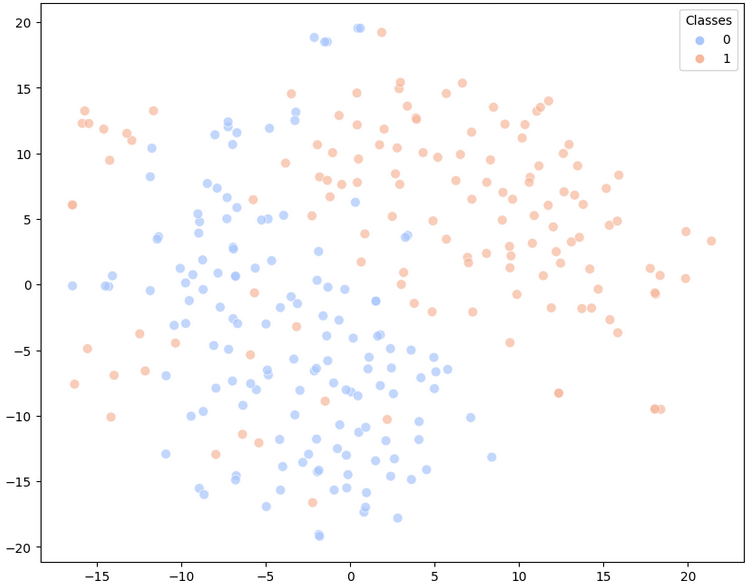}
  \caption{t-SNE of test set \texttt{[CLS]} embeddings (GPTSniffer benchmark).
  \textbf{Left (GPTSniffer baseline)}: The two classes (human=blue, AI=orange)
  are highly interleaved with no clear decision boundary. \textbf{Right (LLMSniffer)}:
  Contrastive fine-tuning produces a compact, isolated AI-generated cluster
  (right) that is linearly separable from the human-written cluster, directly
  explaining the improved classification performance.}
  \label{fig:tsne}
\end{figure*}

Figure~\ref{fig:tsne} shows t-SNE~\citep{vandermaaten2008tsne} projections of
\texttt{[CLS]} embeddings on the GPTSniffer test set for the baseline and
LLMSniffer.

The baseline model produces heavily overlapping clusters for the two classes.
LLMSniffer, by contrast, forms a compact, well-isolated cluster of AI-generated
snippets that is cleanly separable from human-written code. This qualitative
result directly explains the quantitative improvements and confirms that
supervised contrastive learning is the core driver of discriminability.

\subsection{Ablation Study}

Table~\ref{tab:ablation} summarizes the contribution of each component on
GPTSniffer.

\begin{table}[t]
\centering
\small
\setlength{\tabcolsep}{4pt}
\begin{tabular}{lcccc}
\toprule
\textbf{Configuration} & \textbf{Acc.} & \textbf{F1} \\
\midrule
CodeBERT + BCE (baseline)          & 0.70 & 0.68 \\
GraphCodeBERT + SupCon + Linear    & 0.70 & 0.67 \\
GraphCodeBERT + SupCon + MLP       & \textbf{0.78} & \textbf{0.78} \\
\bottomrule
\end{tabular}
\caption{Ablation on GPTSniffer. Upgrading from CodeBERT to GraphCodeBERT with
contrastive learning but keeping a linear head yields no gain; switching to
the MLP classifier (with comment removal) provides the decisive improvement.}
\label{tab:ablation}
\end{table}

The ablation reveals that the gain is not attributable to GraphCodeBERT or
contrastive learning alone. The decisive improvement comes from combining
contrastive pre-training with the MLP classification head and comment removal
preprocessing. We hypothesize that the MLP head is better able to exploit the
non-linear structure of the contrastively-learned embedding space, while comment
removal prevents the model from learning superficial authorship cues in docstrings
and inline comments.

\section{Demo and Deployment}
\label{sec:demo}

We deployed LLMSniffer as a Streamlit web application hosted on Hugging Face
Spaces:\footnote{\url{https://huggingface.co/spaces/mahirlabibdihan/LLMSniffer}}
Users paste a code snippet in any supported language and receive a real-time
binary prediction with a confidence probability. Model checkpoints for both
GPTSniffer and Whodunit are publicly
available.\footnote{\url{https://huggingface.co/mahirlabibdihan/LLMSniffer}}
A video demonstration of the tool is available at \url{https://youtu.be/xkb_vrEfJIY}.
The datasets are hosted at \url{https://huggingface.co/datasets/mahirlabibdihan/GPTSniffer}
and \url{https://huggingface.co/datasets/mahirlabibdihan/Whodunit}. The implementation code is available at \url{https://github.com/mahirlabibdihan/llmsniffer}.

\section{Conclusion}
\label{sec:conclusion}

We presented \textbf{LLMSniffer}, a detection framework for LLM-generated code
that combines GraphCodeBERT with supervised contrastive learning, comment removal
preprocessing, and an MLP classification head. Our ablation study shows that
the full combination is necessary---contrastive learning alone with a linear
head does not improve over the baseline. The complete system achieves
state-of-the-art results on both the GPTSniffer and Whodunit benchmarks, with
the t-SNE visualizations providing a clear geometric explanation for the gains.

Future work will explore: (1)~multi-class attribution distinguishing among
specific LLMs (GPT-4, Claude, Code Llama, etc.); (2)~adversarial robustness
against obfuscation attacks; (3)~integration of AST features alongside
data-flow graphs; and (4)~extension to low-resource programming languages
not covered by current benchmarks.

\section*{Limitations}

LLMSniffer processes snippets up to 512 tokens; longer files require truncation
which may degrade accuracy. The GPTSniffer training and test sets come from
different domains, and while our model handles this well, further domain shifts
(e.g., different LLM families or programming paradigms) may require retraining.
As LLMs produce increasingly human-like code, detection performance may degrade
over time without periodic retraining on updated data distributions. Adversarial
robustness against deliberate paraphrasing or obfuscation remains an open question.

\section*{Ethics Statement}

LLM-generated code detection tools carry potential for misuse. False positives
could unfairly penalize students or developers whose code style resembles
LLM output. We caution against using LLMSniffer as the sole basis for
consequential academic or professional decisions without human review.
We encourage policies that treat model-assisted coding as a tool to be used
responsibly, rather than a practice to be categorically prohibited.

\section*{Acknowledgments}

The authors thank the Bangladesh University of Engineering and Technology for
computational support. We also thank the creators of the GPTSniffer, Whodunit,
CodeSearchNet, and CodeNet datasets for making their resources publicly available.

\bibliography{custom}

\appendix

\section{Dataset Statistics}
\label{app:data}

\begin{table}[h]
\centering
\small
\begin{tabular}{llccc}
\toprule
\textbf{Dataset} & \textbf{Split} & \textbf{Human} & \textbf{AI-Gen.} & \textbf{Total} \\
\midrule
\multirow{2}{*}{GPTSniffer} & Train & 600  & 600  & 1200 \\
                             & Test  & 131  & 142  & 273  \\
\midrule
\multirow{2}{*}{Whodunit}   & Train & 639  & 639  & 1278 \\
                             & Test  & 159  & 159  & 318  \\
\bottomrule
\end{tabular}
\caption{Dataset statistics. Note that the GPTSniffer test set comes from a
completely different domain (source) than the training set.}
\label{tab:datastats}
\end{table}

\section{Hyperparameter Details}
\label{app:hyperparams}

\begin{table}[h]
\centering
\small
\begin{tabular}{lc}
\toprule
\textbf{Hyperparameter} & \textbf{Value} \\
\midrule
Base model            & \texttt{graphcodebert-base} \\
Max sequence length   & 512 \\
Contrastive batch size & 32 \\
Classification batch size & 64 \\
Learning rate         & $2 \times 10^{-5}$ \\
Optimizer             & AdamW \\
Warmup ratio          & 0.10 \\
Contrastive temp.\ $\tau$ & 0.07 \\
Projection dim        & 128 \\
Projection hidden dim & 512 \\
\bottomrule
\end{tabular}
\caption{Hyperparameter settings for LLMSniffer.}
\label{tab:hyperparams}
\end{table}

\end{document}